\documentclass[a4paper]{jpconf}
\usepackage{graphicx}
\usepackage{color}
\begin{document}
\def\d{{\mathrm{d}}}
\title{Projectable Ho\v{r}ava--Lifshitz gravity in a nutshell} 
\author{Silke Weinfurtner$^1$, Thomas P.~Sotiriou$^2$ and Matt Visser$^3$}
\address{$^1$ Department of Physics and Astronomy, University of British Columbia,
6224 Agricultural Road, Vancouver, B.C. V6T 1Z1 Canada; \\
SISSA-International School for Advanced Studies, Via Beirut
2-4, 34014 Trieste, Italy}
\address{$^2$ Department of Applied Mathematics and Theoretical Physics, Centre for Mathematical Sciences, University of Cambridge, Wilberforce Road, Cambridge, CB3 0WA, UK}
\address{$^3$ School of Mathematics, Statistics, and Operations Research,
Victoria University of Wellington, New Zealand}
\ead{$^1$silkiest@gmail.com, $^2$T.Sotiriou@damtp.cam.ac.uk, $^3$Matt.Visser@msor.vuw.ac.nz}
\begin{abstract}
Approximately one year ago Ho\v{r}ava proposed a power-counting renormalizable theory of gravity which abandons local Lorentz invariance. The proposal has been received with growing interest and resulted in various different versions of Ho\v{r}ava--Lifshitz gravity theories, involving a colourful potpourri of new terminology. In this proceedings contribution we first motivate and briefly overview the various different approaches, clarifying their differences and similarities. We then focus on a model referred to as projectable Ho\v{r}ava--Lifshitz gravity and summarize the key results regarding its viability.
\end{abstract}
%
\section{Gravity as quantum field theory\label{Sec:Gravity-QFT}}
%
Even though general relativity has been a very successful classical field theory of gravity, unlike other field theories, {\em e.g.}~electrodynamics, one cannot quantize it in the most straightforward manner, {\em i.e.}~using the canonical quantization or path integral formalism. 
A perturbative loop expansion for gravity results in infinitely many ultraviolet (UV) divergent Feynman diagrams, and at each order the theory requires counterterms of ever-increasing degree in curvature. Consequently, it is necessary to fix an infinite number of free parameters to have a well-defined ultraviolet behavior, and we are left with a quantum theory that has no predictive power at small distance scales, where the counterterms are crucial. In this sense gravity is said to be non-renormalizable. 

There are in principle two different ways to deal with the situation at hand. The first one is to abandon the whole idea of renormalizability, by assuming that gravity is an effective field theory only valid below the cutoff scale. Here the cutoff is a finite, physically meaningful quantity that acts as a short-distance regulator fixing the divergences in the Feynman diagrams. Beyond that scale the theory requires an ultraviolet completion. A typical example of such a theory is the Fermi theory of the weak nuclear force. The other way to proceed is to insist on the concept of a self-sufficient gravitational theory and further investigate perturbative or non-perturbative tools resulting in a renormalizable (or even finite) quantum theory of gravity.

Ho\v{r}ava--Lifshitz gravity~\cite{Horava:2008ih,Horava:2009uw} is grouped within the second category and investigates extensions to Einstein--Hilbert gravity that explicitly break local Lorentz invariance resulting in a power counting renormalizable $d+1$ dimensional theory of quantum gravity. 
In the manuscript at hand key features of Ho\v{r}ava--Lifshitz gravity are reviewed. We then focus on the projectable version of Ho\v{r}ava--Lifshitz gravity~\cite{Horava:2009uw,Sotiriou:2009vn,Sotiriou:2009kx}.

The big appeal of Ho\v{r}ava--Lifshitz gravity as a quantum gravity candidate lies in the fact that --- once the theory is set up --- one has the chance to perform concrete computations and explicitly check its viability. As usual,  it would be very difficult to prove the overall working success of the theory.  Instead, one tends to focus on the various shortcomings in order to eventually establish whether they can be overcome or whether the theory is ruled out. Our intention here is to overview projectable Ho\v{r}ava--Lifshitz gravity and summarize its shortcomings.

%
\section{Emergence of local Lorentz symmetry\label{Sec:LIV}}
%
The key idea behind Ho\v{r}ava--Lifshitz gravity is to sacrifice local Lorentz symmetry in exchange for renormalizability: Local Lorentz symmetry is violated at the non-perturbative level of our field theory, but will hopefully be recovered at long distance\,/\,low energy scales. The particular kind of Lorentz violations we will encounter here have been discussed to some extent within quantum gravity phenomenology~\cite{Mattingly:2005aa,Jacobson:2005aa,Liberati:2006sj}. Let us consider the following class of theories that break Lorentz invariance fundamentally, and that at the kinematical level lead to a dispersion relation which is some function of momentum and mass, $E^2 = m^2 + p^2 \to E^2 = F(p,m)$.
Specifically, the dispersion relations encountered by Ho\v{r}ava--Lifshitz theories are truncated at finite order $2z$,
\begin{equation} \label{Eq:DispRelLIV}
E^2= m^2 + p^2 + (\tilde\eta_{1})_{i} \, p^{i} + (\tilde\eta_{2})_{ij} \, p^{i}p^{j} + \dots + (\tilde\eta_{2 z})_{i \dots j} \, p^{i} \dots  p^{j},
\end{equation}
The truncation coefficient $z$ is also referred to as the degree of anisotropy, adopted from condensed-matter jargon referring to the anisotropic scaling between space and time.

As will be demonstrated shortly, local Lorentz symmetry breaking of the above type can act as a quantum field theory regulator in the ultraviolet for appropriate choices of $z$. Next we focus on two particular examples  and discuss their (power-counting) renormalizability.
%
\section{Renormalizability of anisotropic field theories\label{Sec:Power}}
%
We will lay out in brief a road map to construct  field theories in which Lorentz symmetry breaking acts as a quantum field theory regulator. The first example will be a polynomial interacting scalar field, and the second will be gravity. The study of the scalar field case (Lifshitz scalar) is mainly of pedagogical interest, as any detailed renormalization group calculations for the case of gravity are still pending and all assumptions are based on power-counting~\cite{Anselmi:2008bq,Anselmi:2008bs,Anselmi:2008bt,Anselmi:2008ry,Visser:2009ul,Visser:2009ys}. This is in contrast to scalar field theories in this setting, which have been investigated more closely over the last few months~\cite{Anselmi:2008bq,Anselmi:2008bs,Anselmi:2008bt,Anselmi:2008ry,Visser:2009ul,Iengo:2009ix,Visser:2009ys}.

\subsection{Lifshitz scalar field theories\label{Sec:Anisotropic-Scalar}}
The simplest yet insightful field theory to demonstrate the above concepts  is a (massive) scalar field with polynomial interactions on a flat background,
\begin{equation}
\label{scalaraction}
S = \int   \left\{ \dot \phi^2 - \phi \tilde{D}  \phi - m^2 \phi^2 + \sum_{n=1}^{N} g_n \phi^n \right\} d t \; d^d x ,
\end{equation}
where the differential operator $\tilde{D}$ is given by
\begin{equation}
\label{Deq}
\tilde{D}= -c^2 \triangle + \dots + (-\triangle)^z.
\end{equation}
For a relativistic field theory $z=1$ and $\tilde{D}$ is the standard Laplacian, $\triangle$. These kind of theories are well understood, and we know that $\phi^n$\,/\,$\phi^6$\,/\,$\phi^4$\,/\,$\phi^3$ are renormalizable in $1+1$\,/\,$2+1$\,/\,$3+1$\,/\,$5+1$ dimensions respectively. 

Suppose now that $z$ is left arbitrary and we impose anisotropic scaling between space and time, using the engineering dimensions
\begin{equation}\label{Eq:AS}
\left[ dx \right] = \left[ \kappa \right]^{-1} \quad \mbox{and} \quad
\left[ dt \right] = \left[ \kappa \right]^{-z},
\end{equation}
where $\kappa$ is a placeholder symbol for some object with dimensions of momentum (this scaling is the most convenient for power counting in this case). Choosing the dimensions of $\phi$ such that $[\phi]^2 [dt \, dx^d]=[1]$,
 $\triangle^z$ becomes a marginal operator in eq.~(\ref{Deq}), and the lower-order $\triangle^n$ operators become relevant. 
The coupling constants $g_n$ in action (\ref{scalaraction}) have dimensions $[g_n] = [\kappa]^{d+z- n(d-z)/2} =  [m]^{[d+z- n(d-z)/2]/z}$, which are non-negative as long as $d+z- n(d-z)/ 2 \geq 0$. For $n \to\infty$ this would require simply $z \geq d$. If we apply our experiences from studies of Lorentz invariant scalar theories, non-negative momentum dimensions are a good indication for renormalizability.   In reference~\cite{Visser:2009ul}  the superficial degree of divergence was derived and it was shown that for $z=d$ ($z > d$) and with (without) normal ordering the model at hand is perturbatively ultraviolet finite. 

Before we move on to gravity we would like to stress that although the ultraviolet behaviour appears to be significantly improved by the inclusion of suitable higher order operators in (\ref{scalaraction}) it has already been shown that the one-loop renormalization and evolution of the coupling constants, for two interacting anisotropic scalar fields, is facing the naturalness problem, see reference~\cite{Collins:2004aa,Iengo:2009ix}: The difference in the speed of light between the two fields runs logarithmically to zero, and thus persists at low energy scales. This is in contradiction to our experimental constraints on Lorentz symmetry breaking, unless we impose some unnatural fine tuning on the bare coupling constants.

\subsection{Ho\v{r}ava--Lifshitz gravity \label{Sec:Anisotropic-Gravity}}
The natural setting in which to apply the concepts discussed above to gravity is within the Hamiltonian formulation of general relativity based on the Arnowitt--Deser--Misner (ADM) decomposition of spacetime,
\begin{equation} \label{Eq:ADM}
\mathrm{d} s^2 = - N^2 c^2 \mathrm{d} t^2 + g_{ij}(\mathrm{d} x^i - N^i \mathrm{d} t) (\mathrm{d} x^j - N^j \mathrm{d} t).
\end{equation}
The ADM formalism foliates spacetime into a family of non-intersecting spacelike hypersurfaces, where the metric field $g_{ij}$ is the induced geometry on the spatial slice, the scalar function $N$ is the lapse, and the three-vector $N^i$ is the shift-vector. The extrinsic curvature is defined as 
\begin{equation}
K_{ij}=\frac{1}{2 N} \left\{  -\dot{g}_{ij} + \nabla_{i} N_{j} +  \nabla_{j} N_{i}  \right\}.
\end{equation}
Within this setting one can write down an action of the form
\begin{equation}
S=\int \left[T - V \right] \sqrt{g} \, N \, d^3x \, dt,
\end{equation}
where $T$ denotes the part that contains time derivatives and is given by
\begin{equation} \label{Eq:ExtCurvature}
T (K) = g_{K} \left\{ (K^{ij}K_{ij} - K^2) + \xi K^2 \right\}.
\end{equation}
The potential $V$ on the other hand does not contain time derivative but it can in principle contain any 3-covariant term constructed with the pieces of the ADM metric. We are abandoning Lorentz invariance and therefore we cannot automatically require $\xi$ to vanish, as is the case in general relativity. Similarly, $V$ can now include higher order spatial derivatives instead of just being equal to the 3-dimensional Ricci scalar, as in general relativity.

An action of the above form is clearly not invariant under general diffeomorphisms. It is, however, invariant under the restricted set of diffeomorphisms that preserve the foliation, {\em i.e.}~$t \to t +  \chi^0(t)$, $x^i \to x^i + \chi^i(t,\mathbf{x})$. This symmetry is enough to prevent $N^i$ from appearing in $V$. In the original proposal from Ho\v{r}ava $N$ was also assumed not to enter in $V$ \cite{Horava:2009uw}, however, this constitutes an ad hoc assumption as noticed by \cite{Blas:2009qj} (we will return later to the various variants of the theory).

The way to proceed now is to invoke anisotropic engineering dimensions, as in equation~(\ref{Eq:AS}), and to evaluate the dimensionality of coupling constants of the leading order operators in the UV. Starting from the ADM decomposition~(\ref{Eq:ADM}), we need to have $[N^i]=[\kappa]^{z-1}$ and we are free to choose $[g_{ij}]=[N]=[1]$, resulting in the dimensionality for the line-element $[ds]=[\kappa]^{-1}$, for the speed of light $[c]=[\kappa]^{z-1}$, the volume-element $[dV_{d+1}]=[\kappa]^{-d-z}$, and the extrinsic curvature $[K]=[\kappa]^z$. Under the constraint that the action has to be dimensionless, we obtain for $g_K$ and for the coupling(s) $g_{2z}$ of the operator(s) of order $2z$ in $V$
\begin{equation}
[g_{K}]=[g_{2z}]=[\kappa]^{(d-z)}.
\end{equation}
These operators will dominate at the UV as they are the leading ones in temporal and spatial derivatives respectively. Choosing $z=d$ will make these couplings dimensionless.
(A more detailed treatment of the power-counting and the derivation in general can be found in references~\cite{Sotiriou:2009vn,Sotiriou:2009kx}.)
This implies that by including in $V$ operators of order $2d$, we would end up with a power-counting renormalizable theory, along the lines of the Lifshitz scalar case discussed previously. 

What remains to be discussed is which can be the exact form of $V$. As already mentioned, in his initial proposal Ho\v{r}ava assumed that $V$ depends only on $g_{ij}$ and its spatial derivatives. To simplify things further he proposed a specific combination of terms, dubbed detailed balance (see \cite{Horava:2009uw} for more details on the exact form of V in this case), which is inspired by quantum critical systems and non-equilibrium critical phenomena. The hope seemed to be that such a rigid construction would be imposed as a symmetry, simplifying renormalization group calculations. There is, however, no clear physical motivation for detailed balance apart from simplicity, and instead detailed balance appears to be disfavored by the first consistency tests, see for example~\cite{Sotiriou:2009vn,Charmousis:2009mz,Lu:2009fj}. 

Once detailed balance is abandoned there is no reason to exclude any possible term.
With or without detailed balance, however, Ho\v{r}ava's model propagates an extra scalar mode with respect to general relativity and appears to be burdened with serious shortcomings , such as instabilities, overconstrained evolution and strong coupling at low energies \cite{Charmousis:2009mz,Blas:2009yd,Li:2009bg,Henneaux:2009zb}. In Ref.~\cite{Blas:2009qj} an extension was proposed, after noticing that terms involving $N$ and its spatial derivatives can be included in $V$ without violating the symmetry of the action. The scalar mode in this model does exhibit improved behavior. However, it still possesses strong coupling  at energies orders of magnitude lower than the Planck scale~\cite{Papazoglou:2009fj}. One could do away with the strong coupling if the higher order operators become important at energies below this strong coupling scale, but this requires tuning of their coupling constants (or the introduction of a new scale) \cite{Blas:2009ck,Papazoglou:2009fj}.

We will not review the literature any further here. On what comes next, we will focus instead on a specific version of Ho\v{r}ava--Lifshitz gravity, the so-called projectable version, for reasons that will become clear shortly.
%

\section{Projectable Ho\v{r}ava--Lifshitz gravity\label{Sec:Complete.Horava}}

We have argued that it is indeed possible to construct a power-counting renormalizable theory of gravity by adding ad-hoc spatial derivatives with twice the order of the spatial dimension of the theory. Therefore, in 3 spatial dimensions there is a plethora of terms which, when added to the Einstein--Hilbert action could lead to a theory with the desirable characteristics. As discussed above, there is no real guiding principle that would allow us to exclude any lower or 6th order operators. On the other hand, taking into account all terms is hardly an easy task from a practical perspective. 

Another restriction proposed by Ho\v{r}ava was to impose that the lapse $N$ be just a function of time $t$. The motivation behind this restriction came from the desire to match the reduced symmetry of the action mentioned earlier. There seems to be nothing fundamental in this restriction. On the contrary, it is bound to cause several problems, for instance, it will not allow us to access any gauge choice for which $N$ is not a function of time only.\footnote{On the other hand, it should be noted that the most physically interesting solutions of the Einstein equations, the Schwarzschild, Reissner--Nordstr\"om, Kerr, and Friedmann--Lemaitre--Robertson--Walker spacetimes can by suitable choice of coordinates all be cast into projectable form, in fact with $N=1$~\cite{Wiltshire:2009zz}.} Most importantly, when $N=N(t)$ the Hamiltonian constraint will not be local, as one cannot get rid of the spatial integral after the variation. Taking this shortcomings into account, assuming $N=N(t)$ has also a serious advantage: it allows us to do away with several terms in the action, as all spatial derivatives of $N$ vanish and at the same times many of the possible terms one can write end up differing only by a total divergence. This makes calculations tractable even for the most general action one could write (see also \cite{Sotiriou:2009vn,Sotiriou:2009kx}). In this sense, it is worth exploring this version of the theory, called projectable Ho\v{r}ava--Lifshitz gravity.

Within this setup the most general action can be put together following the lines of Refs.~\cite{Sotiriou:2009vn,Sotiriou:2009kx}:  applying integration by parts (discarding surface terms), commutator and Bianchi identities, and special relations arising in $3$ dimensions (vanishing of the Weyl tensor), one obtains, besides the standard Einstein--Hilbert terms,
\begin{equation}
S_\mathrm{EH} =   \int   \left\{  (K^{ij} K_{ij} - K^2) +  \zeta^4 R  - g_0\, \zeta^6 \right\} \sqrt{g} \; N\; \d^3 x \; \d t,
\end{equation}
new Lorentz-violating terms involving in total eight dimensionless coupling constants $(\xi, g_2, \dots, g_8)$
\begin{eqnarray}
S_\mathrm{LIV} &=&  \int   \Big\{ \xi\, K^2 -     g_2 \,\zeta^2\,R^2 -  g_3 \, \zeta^2\, R_{ij} R^{ij} 
- g_4 \, R^3 - g_5 \, R (R_{ij} R^{ij}) - g_6 \, R^i{}_j R^j{}_k R^k{}_i 
\nonumber\\
&&
\qquad 
- g_7 \, R \nabla^2 R - g_8 \, \nabla_i R_{jk} \, \nabla^i R^{jk}
\Big\} \sqrt{g} \; N\; \d^d x \; \d t. \qquad
\end{eqnarray}
Here we have introduced a scale $\zeta$ in order to make all couplings dimensionless.

Instead of explicitly deriving the constraint and field equations by variation with respect to the lapse (Hamiltonian constraint) and the shift vector (super-momentum constraint), and the induced metric (dynamical equations of motion), we refer the reader to Refs.~\cite{Sotiriou:2009vn,Sotiriou:2009kx}. However, we would like to point out the key difference between the projectable and the non-projectable case. As mentioned previously, in the former the Hamiltonian constraint is not a local one, implying that there is no local notion of energy conservation. At best we can say that the integral of the energy at each spatial slice is conserved. 

It is worth noting that so far we have been working in units that impose anisotropic scaling between space and time simply because this is convenient for power-counting renormalizability arguments. If instead one needs to work in units where the speed of light $c$ is equal to $1$, then this can be easily achieved by setting $dt\to \zeta^{-2} dt$. After writing the action in these new units, see~\cite{Sotiriou:2009vn,Sotiriou:2009kx}, one can read off the Newton and cosmological constant,
\begin{equation}
\frac{1}{16 \pi G_\mathrm{Newton}}= \zeta^2 \quad \qquad \mbox{and}  \quad \qquad \Lambda  = {g_0 \, \zeta^2\over2},
\end{equation}
so that $\zeta$ is identified as the Planck scale.

The main result up to now is that it is possible to include all possible terms and to explore --- within the restricted but complete model ---  the key features of Ho\v{r}ava--Lifshitz gravity. It should be noted that this result can, in principle, be extended to the non-projectable case through a challenging but straightforward calculation.

%
\section{Excitations in projectable Ho\v{r}ava--Lifshitz gravity\label{Sec:Excitations}}
%
One of the most surprising features of general relativity is that, out of the possible fundamental excitations --- spin-$0$, spin-$1$ and spin-$2$ --- only the spin-$2$ excitation, the graviton, is physical. The spin-$1$ and spin-$0$ modes can be gauged away using general covariance. This feature of general relativity is in perfect agreement with observations, which have not so far revealed extra fields mediating gravity, see for example~\cite{Will:2005va}. Therefore, a crucial  viability test for any gravity theory is whether it exhibits other physical degrees of freedom which would have already been detected by observations.\footnote{It is possible to construct condensed matter systems that exhibit spin-$2$ excitations, but in these systems all the lower-spin excitations are also physical due to the lack of general coordinate covariance. See for example Ref.~\cite{Gu:2006yq}.} In $3+1$ dimensions it is difficult to answer this question directly. It is much easier to study the degrees of freedom at each order perturbing around some specific background solution.

Within the framework of projectable Ho\v{r}ava--Lifshitz gravity one can assume that $g_0=0$ for simplicity and  perturb around flat spacetime geometry, along the lines of Ref.~\cite{Sotiriou:2009kx}. After imposing suitable gauge conditions one can show that there exist a (suitably defined) transverse traceless mode ${ \widetilde H}_{ij}$ satisfying the equation
\begin{equation}
\ddot{ \widetilde H}_{ij}   =  -\left[  - \partial^2   + g_3  \zeta^{-2} \partial^4   + g _8  \zeta^{-4} \partial^6 \right] \widetilde H_{ij}.
\end{equation}
Clearly, the spin-$2$ particle has the dispersion relation discussed previously and shown in eq.~(\ref{Eq:DispRelLIV}).

However, there exist an additional scalar mode: the trace $h$ of the perturbation to the metric $h_{ij}$, which satisfies the following equation
\begin{equation}
\label{xineq0scalar}
\left(1-{3\over2}\, \xi\right)   \; \ddot h = - \xi \left\{ -{1\over2}  \partial^2 +  \left(-4 g_2 - {3\over2} g_3 \right) \zeta^{-2} \partial^4 + \left(4 g_7 -{3\over2} g_8\right) \zeta^{-4} \partial^6 \right\} h.
 \end{equation}
 This is a wave equation with a  trans-Bogoliubov dispersion relation.
 At low momentum the group and phase velocity approach the same limiting propagation speed,
\begin{equation}
c_\mathrm{spin-0}^2 =     {\xi \; \over 2-3\xi},
\end{equation}
The low momentum linearized action for this scalar mode is
\begin{equation}
S_\mathrm{spin-0}~=-\int d^3x dt \left[\frac{1}{c_\mathrm{spin-0}^2} \dot{h}^2-(\partial h)^2\right].
\end{equation}
From the last two equations one can infer the following:  for $\xi<0$ and $\xi>2/3$ the scalar modes is classically unstable. On the other hand, for $\xi\in(0, 2/3)$, the mode is classically stable but then the kinetic term has the wrong sign in the action. Therefore, for these values the mode is  quantum mechanically unstable (ghost).  This leaves us with only two special values allowed, $\xi=2/3$ and $\xi=0$.
The latter is the value $\xi$ has in general relativity, whereas the former is an order 1 deviation from this value and we will not consider it futher.
For $\xi=0$ the gauge used to derived the propagator presented above is no longer accessible and we have to do a separate special case study. Using a slightly different gauge led to a peculiar mode growing quadratically with time in Ref.~\cite{Sotiriou:2009kx}. As it was later demonstrated in Ref.~\cite{Wang:2009yz} this mode was actually a gauge artifact and the scalar appears to be frozen around flat background (see also Ref.~\cite{Gao:2009ht} for Friedmann--Lema\^itre--Robertson--Walker spacetimes).

It appears that there is no scalar mode for $\xi=0$ whereas instabilities plagued nearby values. However, there are two different reasons for which setting $\xi=0$ by fiat will not work. Firstly, $\xi$ is a running coupling and we are lacking a symmetry that would allow us to set it to any specific value. Secondly, appearances deceive and the $\xi=0$ choice is not as healthy as it seems. In fact, even though for this value the mode appears to freeze around maximally symmetric backgrounds, this is only due to the fact that it actually gets strongly coupled at all scales \cite{Charmousis:2009mz,Koyama:2009yq}. It seems that we have run out of available options for the value of $\xi$. On the other hand we know that we would have to drive $\xi$ to zero eventually if general relativity is to be recovered at the infrared.

%
\section{Discussion\label{Sec:Conclusions}}
%
After discussing the motivation and giving a brief overview of the various versions of Ho\v{r}ava--Lifshitz gravity, we focussed on one of them, the projectable version where the lapse is assumed to be only a function of time and not space. The main advantage of this version is that it drastically reduces the number of terms that one must consider in the action and it, therefore, makes it possible to perform explicit calculations using the most general action which respects the symmetries of the theory. 

Unfortunately, projectable Ho\v{r}ava--Lifshitz gravity appears to be plagued with serious problems related to the existence of a scalar degree of freedom. This scalar mode turns out to be classically or quantum mechanically unstable in general, and when parameters of the theory are chosen such that the low energy limit corresponds to general relativity, the scalar exhibits strong coupling at all scales. One could still hope for a non-perturbative restoration of the limit to general relativity \`a la Vainshtein in massive gravity \cite{Vainshtein:1972sx}. This is an avenue worth exploring but so far such an effect has not been shown to be present.

In the light of this we could choose to view the projectable version of Ho\v{r}ava--Lifshitz gravity as a useful tool in order to gain insight in the more complete versions of the theory. Indeed,
some of the characteristics of this more limited version, such as the strong coupling, are still shared by other versions, but some are not ({\em e.g.}~the fact that the Hamiltonian constraint is not a local one). Therefore, its study can provide some insight in the other versions, but only once the subtle differences are carefully weighed as well.

To conclude, Ho\v{r}ava--Lifshitz gravity constitutes an interesting quantum gravity theory. It appears challenging to construct a viable model within this framework. In fact there are numerous issues that have not even been systematically considered yet, such as matter coupling, renormalization group calculations, \emph{etc}. Given the fact that sensible renormalizable gravity theories are not easy to construct, Ho\v{r}ava--Lifshitz gravity seems to deserve further study.
%

%
\section*{References\label{Sec:References}}
%

%
\end{document}